A study about who is interested in stock splitting and why:

considering companies, shareholders or managers

Jiaquan (Nicholas) Chen [1] & Marcel Ausloos [1,2,3,#]

[1] School of Business, University of Leicester, Brookfield,
Leicester, LE2 1RQ, United Kingdom

[2] Department of Statistics and Econometrics,
Bucharest University of Economic Studies,
6 Piata Romana, 1st district, Bucharest, 010374 Romania

[3] Physics, GRAPES, rue de la Belle Jardinière 483, B-4031 Liege,
Federation Wallonie-Bruxelles, Belgium

#<marcel.ausloos@uliege.be> corresponding author

## Abstract

There are many misconceptions around stock prices, stock splits, shareholders, investors, and managers behaviour about such information due to a number of confounding factors. This paper tests hypotheses with a selected database, about the question "is stock split attractive for companies?" in another words, "why companies split their stock?, "why managers split their stock?", sometimes for no benefit, and "why shareholders agree with such decisions?". We contribute to the existing knowledge through a discussion of nine events in recent (selectively chosen) years, observing the role of information asymmetries, the returns and traded volumes before and after the event. Therefore, calculating the beta for each sample, it is found that stock splits (i) affect the market and slightly enhance the trading volume in a short-term, (ii) increase the shareholder base for its firm, (iii) have a positive effect on the liquidity of the market. We concur that stock split announcements can reduce the level of information asymmetric. Investors readjust their beliefs in the firm, although most of the firms are mispriced in the stock split year.

## 1. Introduction

Many financial decisions taken by managers seem to have no benefit to the firm at all. In particular, instead of profit, stock split seems to bring more cost. Stock split is an action in which the firm decides to divide its shares pro rata. Theoretically, after shares are distributed, the price per share should also drop pro rata. Moreover, because the number of shares increases, the firm needs to pay extra dividend to those additional shares. Yet, if a firm would like to have more shares in the stock market, it could "simply" issue new shares to the market. By that, the firm could also gain in equity and cash. Thus, the stock split is not an occasional decision, but appears to be paradoxical.

There are many studies related to this topic, and many hypotheses were built; let us quote Lamoureux and Poon (1987), Lakonishok and Lev (1987), Ikenberry et al. (1996), who can be considered as the pioneers. Our paper has a more modern approach: we test hypotheses with a selected (as explained below) database, to answer the first question "is stock split attractive for companies?"; in another words, "why companies split their stock?" But in this paper, therefore, we also report our study on 'stock splits' decision in order also to try to answer the questions "why managers split their stock for no benefit?", and "why shareholders agree with such decisions?".

## 2. Literature review

As mentioned in the Introduction section, extant literature has discussed stock splitting causes and consequences. However, many factors are confounding. Through a bootstrapping approach, we have decided to investigate only a few variables which might be in our opinion the most relevant ones to our query about causes and consequences of stock splitting. In line with Easley et al. (2001) microstructure approach, we have chosen to investigate (i) whether the stock splitting decision much increases the number of stockholders, in particular small investors, (ii) whether stock splitting truly decreases the information asymmetry on firm value between investors and firm managers, and (iii) whether the event improves the liquidity of the market, i.e, wondering why managers do not simply issue new shares on the market. Thus, the following literature review concentrates on these three hypotheses. We are aware that there are many other potentially interesting hypotheses.

### 2.1 On Hypothesis 1: To attract more shareholders.

Baker and Gallaher (1980) sent a questionnaire to chief financial officers, of two groups of New York Stock Exchange-listed firms. Each group had 100 members, one who had the history of issuing a stock split during 1978; the other group had not such a history. Officers highly responded that actual split decision contributes to attracting more shareholders. From the e-questionnaires, four statements can be globally extracted: Stock splits appear

1. to make more convenient for small stockholders to purchase round lots
2. to keep the firm's stock price within an optimal price range
3. to increase the number of shareholders
4. to make stocks more attractive to new investors or speculators.

They further referred that those managers who made stock split decisions were taking more care with small investors than with institutional investors.

Easley et al. (2001) referred to Copeland's (1979) work, trading range hypothesis:

> *"The clientele preferring a lower price range in usually thought to be uninformed or small investor."*

Nevertheless, one may wonder why managers want to attract small investors.

One of the explanations from the financial executives' side is the value of a share and market confidence (Baker & Gallagher, 1980). Financial executive's expectation is to lower the percentage of shares that are held by institutional holders, for example, investment companies. Because of the high percentage holding by few institutes, these take a large part of trading in the market and could cause some unwilling-to-see influences on firms. Institutional investors which hold a large number of shares could empty their stock and have cash returns in a very short period. However, a reasonable behaviour could be a simple movement; it also could be complicated. Any investor main purpose is to maximize profits; when a renowned investor throws for sale shares it owns in a company, it gives a signal "*I believe this is where I can access the maximum profit, I can have from this stock and now am out*". The movement brings price reduction and public's alertness, whence potential small investors.

Another explanation on the managers' side is the right of control and different risk attitude (Powell & Baker, 1993/1994). The company is always run by managers but is in fact held by a few top managers and shareholders. Although shareholders do not have right to control day-to-day operation, shareholders can assess their right at shareholders' meeting to vote for the important movement. The result of voting sometimes will be against managers' will, because of differences of aims between managers and shareholders, and risk attitude between owners and investors. When the company grows, it requires sustention for expansion from the market. The top manager of the company can either do selling shares or taking loan, or both. In the shared ownership businesses, the owner could explore the shareholder base to have a larger diffusion level of control right, to avoid takeover threat by major shareholders.

Arguments as here above might give some sense about why companies or the top manager want to increase their shareholder base, but Grinblatt et al. (1984) wonder if it is possible that this split announcement has any effect in attracting small investors. Grinblatt et al. (1984) examined daily stock price returns on days around the announcement date and compared them with the following benchmark period over forty trading days, from day 4 to day 43. The data reported that the mean two-day returns of days around announcement date for a split and stock dividend sample of 1762 firms was 3.41%; the mean two-day return for a forty-trading-days benchmark period was only 0.10%. Grinblatt et al. (1984) also found an alike price response in the sample of 244 pure stock split announcements and 84 pure stock dividend announcements. The mean two-day return on a stock split announcement was 5.87% and for a stock dividend announcement was 3.29%. Comparing those with the mean two-day return for the subsequent forty days, it was 0.16% for the stock split and 0.14% for the stock dividend.

### 2.2 On Hypothesis 2: To level off information asymmetries on firm value

Precise knowledge about a firm financial status is usually differently existing between the managers and investors. Yet, such an information is important to shareholders and investors. They can estimate the business value only from information about each decision taken by the firm. However, too much trust and limited assessable information, both make a pyramid scheme. To avoid a Ponzi-like scheme, having access to information is the way to protect the investment. However, most internal users have deeper and more updated information than external users. If managers believe that the value of outstanding shares is less than it should be, managers may act to influence the market. However, if the management publishes an stock split announcement, the stock might be subsequently mispriced by the market. In fact, the manager should indirectly call attention from markets or analysts to analyze the fundamental value of the underlying stock; any result coming out from the independent party would be much more trustable and acceptable by investors, - the same logic holds with the auditor in a financial reporting. Therefore, stock split announcement can be an invitation to the public to re-evaluate the firm attraction. Thus, stock split also adjusts, or is expected to reduce, the level of information asymmetries.

Nevertheless, let it be pointed out that Easley et al. (2001) ''*do not find evidence consistent with the hypothesis that stock splits reduce information asymmetries*''. This was re-examined by Karim and Sarkar (2016) who also wonder if stock split is the signal of undervaluation. They sampled ordinary common stocks from the ~~CRSP~~ Center for Research in Security Prices (CRSP) database, from 1973 to 2014 (https://www.crsp.org/). The analysis shows that over half of the stock splitting firms had been mispricing at surrounding split announcement dates; a significantly larger percentage of stock splitting firms were overvalued as compared to non-stock splitting firms. One reads in Karim and Sarkar's (2016) that 81.9% of stock splitting firms were mispriced in the stock split year. However, about 50% of non-stock splitting firms also faced a mispriced situation. Thus, Karim and Sarkar (2016) evidence that the information known by internal and external investors is quite different.

On the contrary, if stock splitting firms are overvalued, Grinblatt et al. (1984) argued that managers may think about the trading range of their stock in light of the optimal trading range hypothesis. Guo et al. (2008) studying the Tokyo Stock Exchange support the hypothesis which predicts that the reduction in price due to the splits later on attracts the usually less well-informed small investors, - a behaviour considered to be due to some reduction of the information asymmetry.

Brennan and Copeland (1988) find that stock split is a signal that firms give their internal information about future earnings. In the case that pre-split price and firm size are under-controlled, split factors increase the difference between actual and predicted earning. The authors believe that if the signals of the stock split are advising about the company's future earnings, then a readjustment of investor beliefs at the announcement date should correspond to the firm's earnings forecast error. Brennan and Copeland (1988) used samples that had stock dividends and splits history during 1976-1983 (because machine-readable earning forecast data were available to access); 225 of those samples are "SD/SS" (stock dividend/stock split) announcements only" samples. The result from the two groups is different, after ruling out the influences of others announcement, the SD/SS group gave a more reliable answer: the coefficient on FE (One-year-ahead percent earnings forecast error) and *uspfac* (unexplained split factor) have *t*-values of 2.11 and 2.14 with probability values less than 0.02. The data refers that investors readjust their beliefs about the value of the firm at announcement date in a manner that corresponds with the firm's following abnormal return. According to Brennan and Copeland's (1988) study, there is a strong relationship between split factor signals and announcement returns. They suggested SD/SS announcement has an effect on the future earning, and more importantly that investors do react upon. A good or bad signal to future earning depends on the firm's choice, but investors respond to it; yet, not all with the same logics since the return distribution is both skewed, peaked, and with non-Gaussian tails (Dhesi and Ausloos,

2016).

*2.3 On Hypothesis 3: To provide stock splits/dividends, new shares, and liquidity.*

If firms would like to increase the number of shares in the market, one wonders why managers do not just issue new shares to the market, - since issuing new shares likely could raise money for the firm and increase the capital of the firm (Yagüe, Gómez-Sala, & Poveda-Fuentes, 2009).

In its ‘’tools and guidance for businesses”, the UK government (GOV.UK, n.d.) states that those firms which want to issue shares must inform Companies House, an Agency of the Department for Business, Energy & Industrial Strategy. For some firms, the manager may need a special resolution (75% or even 95% agreement from all shareholders) to shift firm’s share structure. The manager must tell Companies House within 15 days if the manager wants to issue more shares of the company. The most difficult requirement to achieve is ‘special resolution’ because it needs a majority of shareholders to agree. It will be even harder if the decision cut shareholders” benefit – issuing new shares. By using the supply-demand curve, when there is a sudden increase in supply without others change, it causes over-supply; thus, the price will go down and shareholders’ wealth go down as well. Also, issuing new shares to the capital market will attenuate the percentage shareholding for every existing shareholder. The percentage shareholding affects the control rights and necessarily attract more people to share the surplus. In this respect, shareholders need to purchase extra shares to keep the same level of control as before.

Considering those disadvantages to shareholders in issuing new shares, it is reasonable that management seeks alternative options, like stock split and stock dividend. If management wants to split the company’s stock or increase the stock dividend, it also requires a special resolution support. However, stock split and stock dividend provide a reason to allow current shareholders to have a concurring vote. Indeed, the stock dividend increases the current dividend which will directly increase shareholders’ income, and once the dividend per share is increasing, the price per share will be raised as well, based on dividend discount model. The stock split increases the number of shares in the market. However, after stock split is finished, every shareholder still has the same control right of the firm as before. Shareholders could sell additional shares as their will is to make cash income. Therefore, the decision would be passed easily because it benefits all shareholders.

Furthermore, stock splits and stock dividends have benefit for the firm: stock splits and stock dividends are unlike most cash dividend, they do not have a direct effect on the future cash flow of the firm (Grinblatt et al., 1984). Also, since investors do not need dividends to convert shares to cash, they will not pay higher prices for firms with higher dividend pay-out, in words, dividend policy will have no impact on the value of the firm (Modigliani, 1961). Thus, stock split and stock dividend have not only similar effects, but their principles are very similar as well. Grinblatt et al. (1984) wrote that

> *"... All 'stock dividends' exceeding 25% are treated as splits and do not affect retained earnings, ... stock distributions between 20% and 25%, ... are usually treated as stock dividends."*

Stock splits and stock dividends both could be represented as an increase of total stock dividend. Shareholders will receive an extra dividend from either additional share or dividend increase announcement. Also, based on the dividend discount model (Wiiliams, 1938), the stock price will grow with increasing of the dividend. For the stock dividend, it is hard to avoid but not for the stock split. A higher price is very attractive for shareholders to sell their shares, by that the price per share will drop due to increase in supply. However, in the Baker and Gallagher's study (1980), major chief financial officers have an expectation of low-price range rather than high. The idea (Brennan & Copeland, 1988) stands from folklore that a split will increase the demand for a stock among small investors, which will, in turn, improve the stock's liquidity, and hence presumably raise its price. In Brennan and Copeland's writing (1988), stock splits also cause a price increase, but the difference is that the price after stock split will be set on split-adjusted basis. It means the more additional shares are issued to shareholders, the lower price it will be. Even if there is an increase in price, the price per share will still be lower than the price before the split.

This also has been suggested in McNichols and Dravid's review (1990): a lower range of share prices enhances the liquidity of a firm's shares. In fact, Barker (1956) had found that the average gain in common stock shareholders for split-up stocks was 30% whereas 6% for companies whose stocks had not been split. Copeland (1979) explained that an individual who holds one share and likely to sell it to one buyer, may sell to two people after the two-for-one split. Since more shares need to be sold in the market, it potentially increases the number of shareholders, trading volume and liquidity. Liquidity is important to all investors who want to get into the market or get out, because liquidity measures how convenient for investors to transfer shares to cash. High liquidity means that there are many traders active in the market. They could place their order, could exchange information, could bargain with others, make an agreement, etc. Traders are reacting to the market. For those who want to join in or leave, high liquidity gives them an effective platform to publicize their will, which their information will be noticed and will be read frequently. Liquidity will affect the gap between buying and sell: all traders in the market, both buy side and sell side, have access to mass information about who want to sell or purchase how many shares in which price. All traders want to gain the maximum benefit from each transaction, so they try to sell at a high price and buy at a low price. This is the bid-ask spread notion.

However, purposes of investors are different; some may want to purchase shares and hold it to be a shareholder, to have some control right on the firm, while others may treat it as an investment. Moreover, investors would like to minimise their cost but according to different aims: the acceptable price for each investor is independent and multifaceted; some would like to pay more, some do not. In the market, the buyer and the seller who want to trade oppositely at the same price will reach a deal; their order in the order book will be executed, and then the next order fills in the blank. It is a reasonable assumption that people feel stressful when they watch that mass orders are over and new orders fill in, but nobody trades them. Thus, the investors might reduce their will and let the price of the order be closer to the recently dealing range in order to make their order more interesting and worth to be considered by others.

## 3. Methodology

Thus, from the literature review, it seems that we can indeed focus on three possible reasons ("hypotheses") for top managers to split a firm stock:

H1. To increase the number of shareholders.

H2. To reduce the information asymmetry.

H3. To improve the liquidity.

To test whether those three hypotheses have an actual effect on the decision of stock split, the best way seems to find out whether there are financial changes which match those hypotheses. In other words, it is worth to investigate whether the stock price and the operating performance reflect the change, if there is a change.

### *3.1 The research method for Hypothesis 1*

It appears costly and time-consuming to verify Hypothesis 1 directly because there is not a document to clarify it in detail but only an approximate number of holders of each class of common equity of the registrant at the latest practicable date. (Cornell Law School, n.d.) (U.S. SECURITIES AND EXCHANGE COMMISSION, 2016). Also, only an approximate number of shareholders existence is found in U.S. companies' annual reports. Thus, under reasonable assumptions, we measure the change in the number of shareholders indirectly, that means through public data information.

The first objective of Hypothesis 1 is to increase the number of shareholders. To increase the number of shareholders, new investors must purchase shares to become a shareholder. Therefore, the amount of trading should increase. We compare the trading volume during 30 days before and after the stock split date, to see if there is any effect in the trading volume. The amount itself does not matter in this research but the percentage change in trading volume and the trend line of trading volume are of interest. Indeed, the trend line is a tool to predict how the investors trade in the market, their attitudes to shares – "are investors more willing to be holding shares to receive dividends or selling them for cash?" Also, the slope of trend lines measures growth rates of total trading volume (Lo &Wang, 2000) before and after the division.

Comparing slope rates before and after the split, the difference between them also reflects a change of attitude. A low slope means that investors are gradually reluctant to exchange shares, whereas a high slope means that investors have an open mind to trading orders. A significant increase in trading volume could be a sign that more people exchange their shares to others, in another word, distributing shares.

On the other hand, companies reduce the price of the share by the stock split in order to attract new shareholders (Baker & Gallagher, 1980). Indeed, some investors are price sensitive (Mohr & Webb, 2005), a low price could draw more attention. So, price dropping is the second signal which a manager wants to see. To verify Baker and Gallagher~~’s~~ (1980) finding, historical stock data after the stock split does show how the market responded to stock split announcements. We did gather historical stock data about 183 days, from -91 days to 91 days. (Day 0 is the split day, -1-day mark one day before the split day and 1 day mark one day after the split). Those 183 days (six months) are divided into three groups, every 91 days (two months) being a group. The role of the first group is to give a “normally expected” value of the stock because the first group is three months away (ahead or after) from the split day. The second group shows how the market reacts to the share diversion. There are three expected responses: decreasing, unchanged or increasing. The third group is to measure the effect of the stock split is a short-term effect or a long-term effect.

### *3.2 The research method for Hypothesis 2*

Information asymmetry is hard to measure because external users do not know the daily operation; indeed, it is hard to know how well a firm is operating and how many information internal users cover. Thus, the published information is the only way to measure information asymmetry. Information asymmetry affects the public’s attitude to the firm indeed, as well demonstrated still recently (Romito & Vurro, 2021).

In that line of reasoning, managers can use stock split announcements to attract the public’s attention. It makes analysts revalue firms’ future cash flow. If revaluation is positive, it means that the firm is better than what the public expects. The result of revaluation not only re-adjusts the public's attitude to the firm but also enhances the confidence level. Hence, investors are willing to pay more to become a shareholder to share the profit. This is marked in the return distribution (Dhesi and Ausloos, 2016). The more money that investors are willing to pay, the higher value the firm has. If there is such a ''positive information asymmetry", the firm should be better than before, or can do better in the future.

Also, the change of price also helps to test whether Hypothesis 2 reflects that the firm is overvalued or undervalued. If the firm is undervalued, this means that the market value of the firm is lower than what it is supposed to be. Thus, the price in the third group should be over the price in the first group. On the contrary, if the firm is overvalued, the price in the third group should be below the price in the first group. Thus, we use the same methodology as for Hypothesis 1, but with a longer time span, – one year, to measure the change in the price, to ensure that the change of price is relatively permanent (Ausloos and Ivanova, 2003; Ferreira and Dionísio, 2016).

One can also measure the performance of the firm through data in its annual reports. Many indicators are used to measure performance, but we only use the data of the net profit and ROE (return on equity). Because the extra benefit which shareholders could gain is depended on the yearly earnings; the ROE ratio shows how well the firm uses equity holders' assets. A high ratio reflects that the firm is better in its responsibility – maximise equity holders' wealth. If the change of ROE ratio is followed by a change in stock price, then it might mean that the public agrees with what the firm believed; otherwise, changes will be the opposite. Also, the announcement might be a sign of future earnings. Brennan and Copeland's (1988) found that investors readjust their beliefs about the value of the firm at announcement date in a manner that corresponds with the firm's following so called abnormal return. The abnormal return comes from trading shares in the market. The factor that leads to abnormal return is an unexpected increase in the stock price. The sudden increase could be provoked by the improvement in operating performance, - but also could be generated by the political and economic environment.

Thus, we measure the increase ratio of daily stock price, and compare it with the 120$^{th}$ historical stock price before the split date. These percentages indicate how much profit (in per cents) investors could receive if they held shares for that long. However, there will be an expectation for the growth speed: if the stock price follows the expectation, then there is no abnormal return. Thus, an unexpected growth rate can generate an abnormal return. For this observation, we use data over 120 days before the split date, and calculate returns one, two, three and four months after the split date. Because Brennan and Copeland (1988) suggested that investors readjust their beliefs at announcement date, we use the average price between day 59 and day 60 as the virtual demarcation day for calculation. The result represents after adjustment, whether investors have a strong idea about the firm's market value and future operation.

1. Steps for calculating the return.
1.1. Identify the adjusted closing price on the first and last day of the period before the split.
1.2. Divide the adjusted closing price at the end of the period by the one at the beginning of the period. This gives the "normal return rate".
1.3. Then, find the index's adjusted closing price on the first and last day of the period after the split.
1.4. Divide the ending adjusted closing price by the beginning adjusted closing price.
1.5. Multiply the stock's beta. This gives the return rate after the split ~~with the~~ has influenced ~~by~~ the market.
1.6. Subtract the two results; the difference value shows whether the stock generates better returns than expected.

2. Steps for calculating beta.
2.1. Calculate the percent change period to period for both the stock price rate ($r_{sf}$) and the risk-free rate ($r_f$).
2.2. Find the Variance of the stock price.
2.3. Find the Covariance of the stock price to the risk-free rate.

2.4. Using the formula to calculate beta: $beta = \frac{Corr(r_f, r_{sp})}{Var(r_{sp})}$.

Comparing data obtained at step 1.2 and 1.5 allows the results mathematically to prove that the firm response to investors is within their expected aim.

### 3.3. The research method for Hypothesis 3

Liquidity measures how easy the transfer is between cash and securities. Large trading volume could represent high liquidity, but trading volume is sensitive to announcements and news (Lo & Wang, 2000). Any decision or news report could cause panic selling or buying and then caused extraordinarily high trading volume, which is a panic effect. Stock split announcements might be one of many events that can create the panic effect. To avoid (spurious) panic effect, we only test the trading volume for period 1 – 90 days before the split date; 90 days is thought to be a quite extended period; it cannot necessarily remove the panic effect but is expected to minimise it. Also, the time period is the same as that used for the price gap. The trend of Period 1 gives a view on what the market thinks about the sample company without being affected by the stock split. We test the trend for period two – a year with the split date as the middle day. The trend of period 2 will give the same view as with the effect of the stock split. Also, comparing test 1 (test for period 1) and test 2 (test for period 2), we can observe whether the liquidity increases or not, and how long the change of liquidity appears to remain. If the stock is a high liquidity type, then every trader within the stock can execute the order with a pleased and fair price. Of course, we repeat, buyers want to buy at the lowest possible price while sellers wish to sell at the highest possible price. The result is that investors trade with the market price. There is a possibility for a price gap to exist (the difference between high and low) because the current supply could either over or under fit the demand. But, due to the significant number of supply and demand, the price gap should not be significant.

Thus, we not only test the change in trading volume after the stock split but also the price gap. The price gap is supposed to be smaller than that before the split if the liquidity is increasing, as found in McNichols and Dravid's review (1990).

Theoretically, the price gap should be dropping after the split, but the gap is not expected to drop forever. The drop of the gap is a corresponding consequence of the increase in the liquidity, but the gap could also have some slight repercussion, as the result of the increase in trading volume and opportunists.

## 4. Data selection and description

We have chosen some data to be examined such that a small time interval occurs when a stock splitting event happened, in recent years, with some further constraint that there is some possibility to center the pertinent time interval in a wider time interval of approximately equal size before and after the event. Moreover, we have decided to choose the interval such that there are enough data points for drawing random cases, - with not too few events in order to have some meaningful statistics, but also not too many cases, in order not to introduce spurious firm heterogeneity which would have blurred the conclusions.

Nevertheless, in the stock split calendar, which is offered by *investing.com*, (https://www.investing.com/), there were 7727 events in 2013 and 2014, - too many for a study partially demanding some qualitative understanding. Besides, although there are more stock split events before 2013 and after 2014, those events are also too old to find the information of the day, or too young to estimate the impact. To remain unbiased, and for a reasonable statistical analysis, we use the random code in Excel to pick 13 numbers from 1 to 7727. Every number points out to a "stock split event", including information on a company, date of split and ratio of the split. However, four samples could not be used and were moved out from the sample list because

1. The language of the annual report is not English.
2. There are different standard**s** in annual reports
3. There is no historical yearly report documents remaining on the homepage
4. We were incapable of collecting full data of interest.

Thus, there were only nine samples remaining for the investigation. For every pattern, we did collect available historical stock data, starting from 1st Jan 2013, and the operating performance for two years after the split year. The historical stock data is collected from *investing.com*, and any data related to statements are obtained from companies' annual financial report. The vast range of historical information is preparing for horizontal comparison. The parallel comparison to others allows us to affirm the effect of stock split announcement to the company as being relevant of investigation.

For completeness, let us report the split ratio for the nine samples in Table 1.

Table 1. Split ratio (potential trading number of shares *after* the split exchanged for each single share *before* the split) for the nine examined samples.

| #1 | #2 | #3 | #4 | #5 | #6 | #7 | #8 | #9 |
|---|---|---|---|---|---|---|---|---|
| *1.25* | *1.1* | *1.015* | *1.068* | *1.569* | *2* | *1.333* | *1.011* | *4.899* |

## 5. Results

### *5.1. Hypothesis 1*

Consider H1, the number of shareholders and their possible trade volume. Figure 1 displays the total trading volume of the 9 samples over 30 days before and after the split. Four samples (sample 1, 4, 6 and 7) have a noticeable increase (around 10% or above) in trading ~~times~~ volume after the split, markedly for sample 6 and sample 7; for one sample (sample 8) the volume remains stable; the others (sample 2, 3, 5, 9) have dropped in trading volume, markedly for sample 5. One could conclude that the stock split event does not have a~~n~~ systematic effect on the trading volume market. However, some difference can be noticed in the total trading volume between before and after the split, as observed from Figure 1, and requests some further attention.

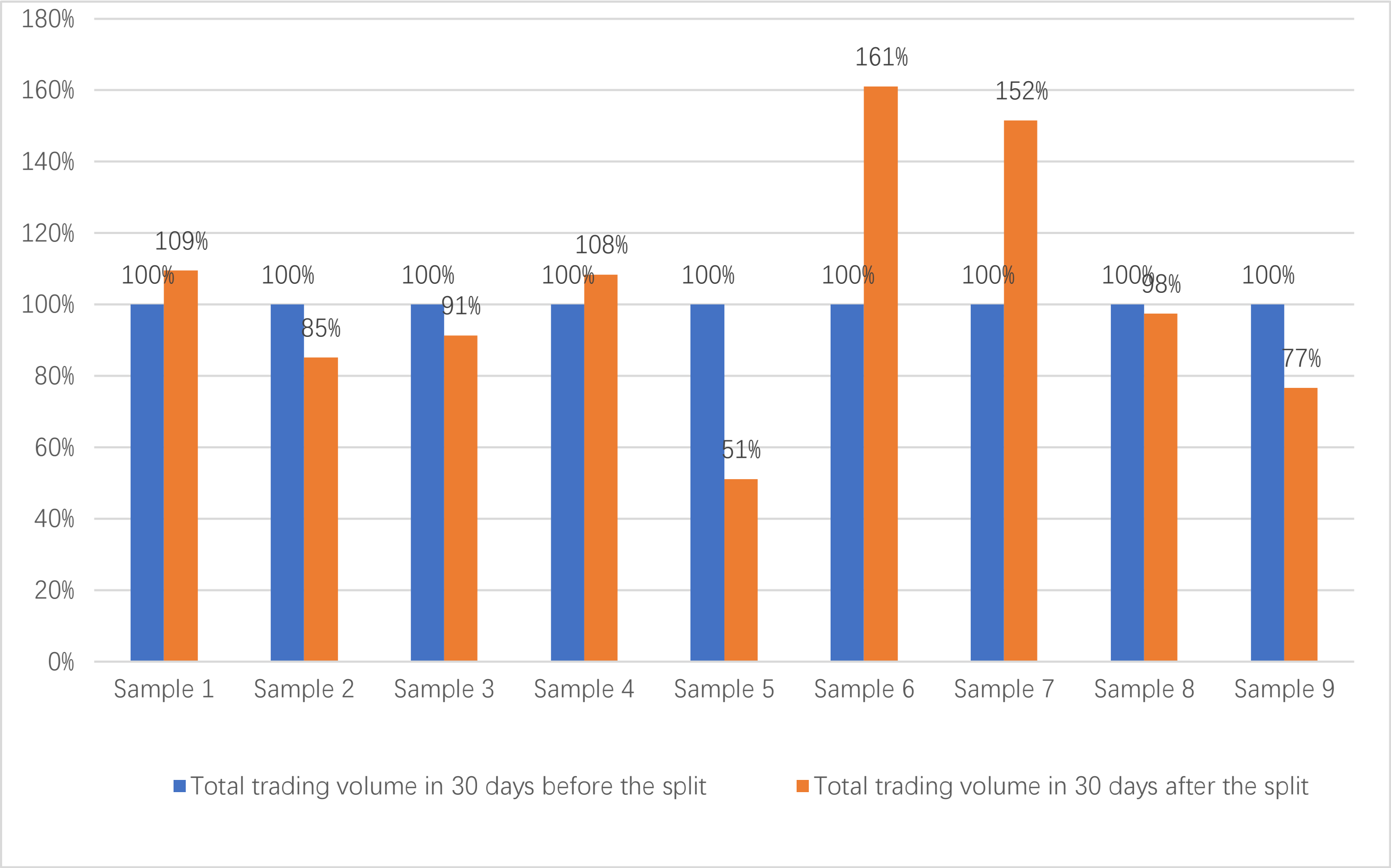

Figure 1. The total trading volume in 30 days before (left column; blue online) and after (right column; orange online) the split. The total trading volume before the split is treated as benchmark (100%) in order to check the change in the total trading volume after the split.

Figure 2 displays a macroscopic approach view of the nine split events. Recall that the samples are randomly selected from all companies that have a history of the stock split between 2013 and 2014, without any category limited constraint. Thus, those nine samples can be considered to give a mini view of the entire "stock split market". Figure 2 demonstrates that the total trading volume after the split had a slight increase, 4%. However, in this time span, we are not considering an increase in a single stock case, but a whole stock split market, with millions trading per day. This 4% increase of total trading volume in fact is a massive impact on the market. Thus, one concludes that stock split events much affect the market, in particular, stock split announcements enhance the number of trades.

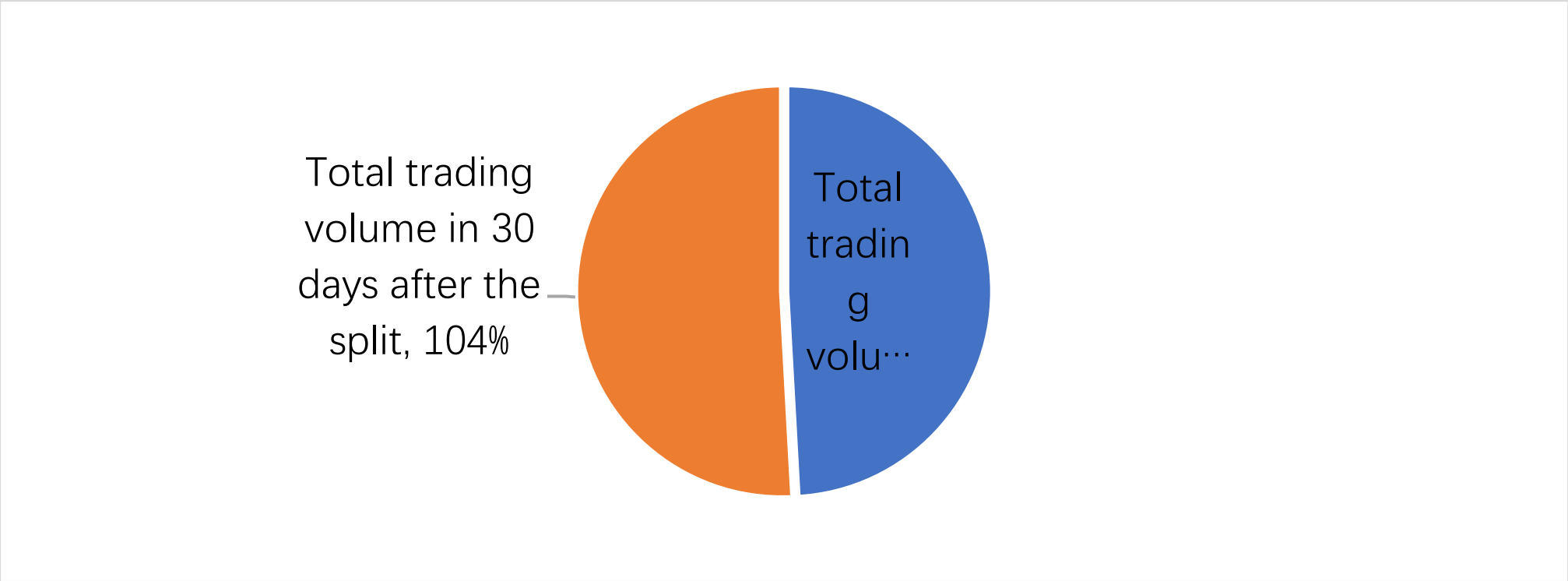

Figure 2. Total trading volume over 30 days after (orange area online; left side of the cake) and before (blue area online; right side of the cake) the split.

More specifically examining the daily exchange, Figure 3 presents the daily trading volume 30 days before and after the split date separated; Figure 4 provides slopes for trend lines. From Figure 3, one observes irregular and significant sizes change. The market was reacting to the split announcement. Notice from Figure 3 that over half of the samples reached the peak of those 62 days trading after the split date. Consider the difference before and after – the extra shares contributed, it seems that investors are more willing to distribute their shares to others after the split. But, from Figure 2, one observes that the total trading volumes before and after the split are almost unchanged for this case selection. Thus, the trading amount after the split is not significantly different from the amount before, but nevertheless is "more concentrated". This concentrated-trading situation implies that there was an explosive demand for shares.

Since the stock split does not have a significant effect on each event total trading volume, and the global measure, we turn our attention to trend lines. From the table in Figure 4, it is seen that samples often have an adverse change in slope rates. In another word, investors' aspiration of trading after the split day is weak; they prefer to trade their shares before the split date. However, there are some remarkable positive numbers: 12172% and 4449%. Those (incredibly) high numbers point out toward a probability that for minority stocks, trading will be more frequent after the split date. To verify this probability, we can combine Figure 4 and Figure 1. From Figure 1, sample 7 had a 52% increase in the total trading volume after the split, but sample 9 had a 23% decrease; consequently, one deduces that there is no indication of direct connection between slope rates and total trading amounts. Thus, in the short term, stock splits affect the trading volume, but in most of the samples' cases, the effect is starting to disappear after the split date.

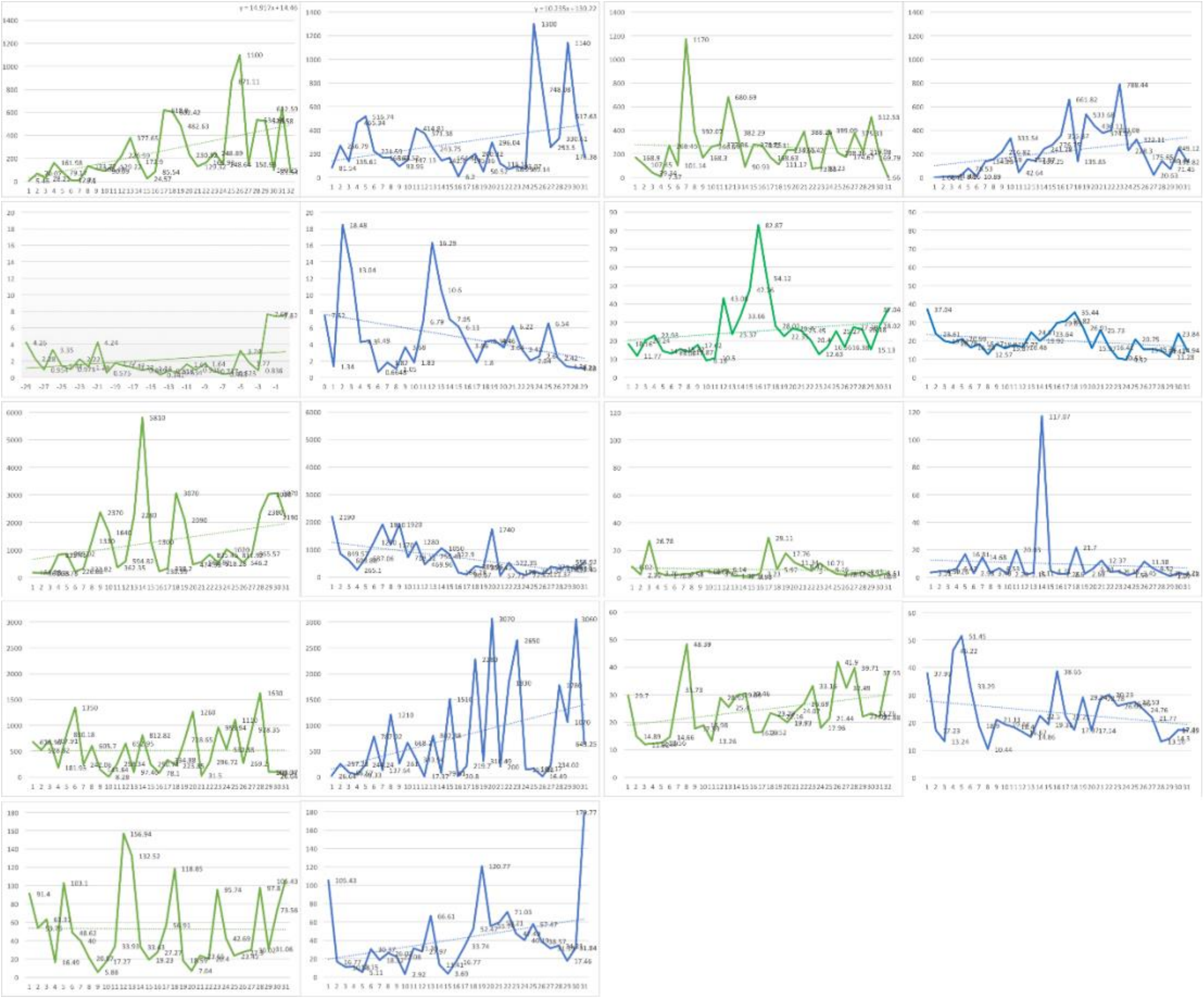

Figure 3. Daily trading volume over 30 days before and after the split date for nine samples, from sample 1 to sample 9, from left to right, top to bottom. Each left panel (green line online) represents the daily trading volume starting 30 days before the split. Each right panel (blue line online) represents the daily trading volume up to 30 days after the split. A linear fit is reported in order to suggest the trend behaviour; see also Fig. 4.

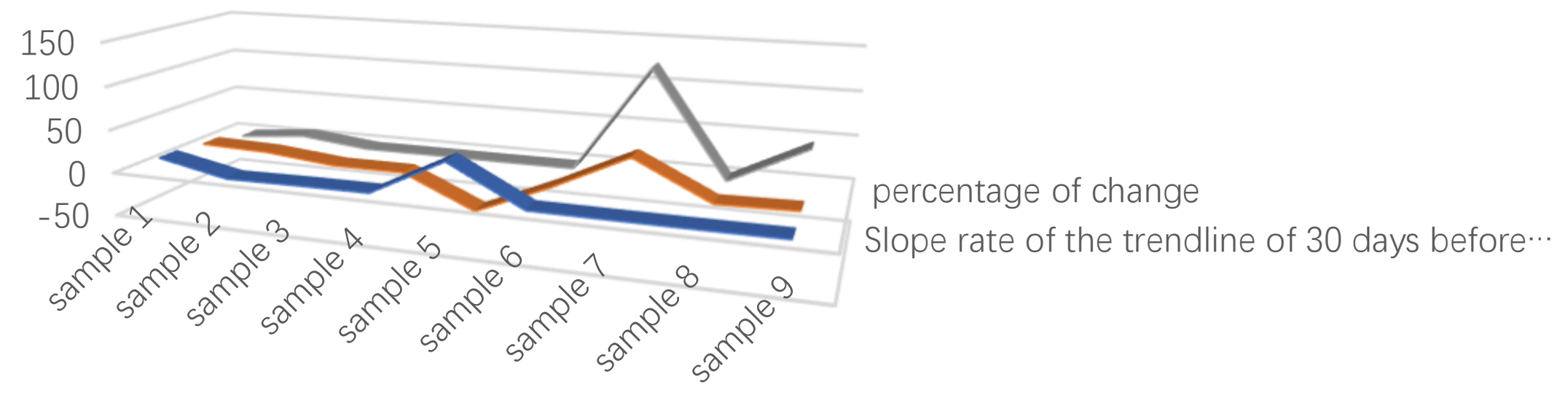

| | sample 1 | sample 2 | sample 3 | sample 4 | sample 5 | sample 6 | sample 7 | sample 8 | sample 9 |
|---|---|---|---|---|---|---|---|---|---|
| Slope rate of the trendline of 30 days before the split | 14.917 | -1.2535 | -0.0684 | 0.3251 | 43.23 | -0.0733 | -0.3435 | -0.3553 | -0.0337 |
| Slope rate of the trendline of 30 days after the split | 10.235 | 7.927 | -0.1833 | -0.2149 | -36.822 | -0.1937 | 41.468 | -0.2836 | 1.4657 |
| percentage of change | -31% | 732% | -168% | -166% | -185% | -164% | 12172% | 20% | 4449% |

Figure 4. Slopes for trendlines over 30 days.

Let us now consider the specific aspect of the hypothesis concerned about reducing the price, through stock split, in order to attract new investors. If this strategy is working, then the price before the split date should be higher than the price after the split. However, Figure 5 shows a different pattern. The red lines – the stock price without being affected by the extra shares, is not obviously above other lines, but is intertwined with others.

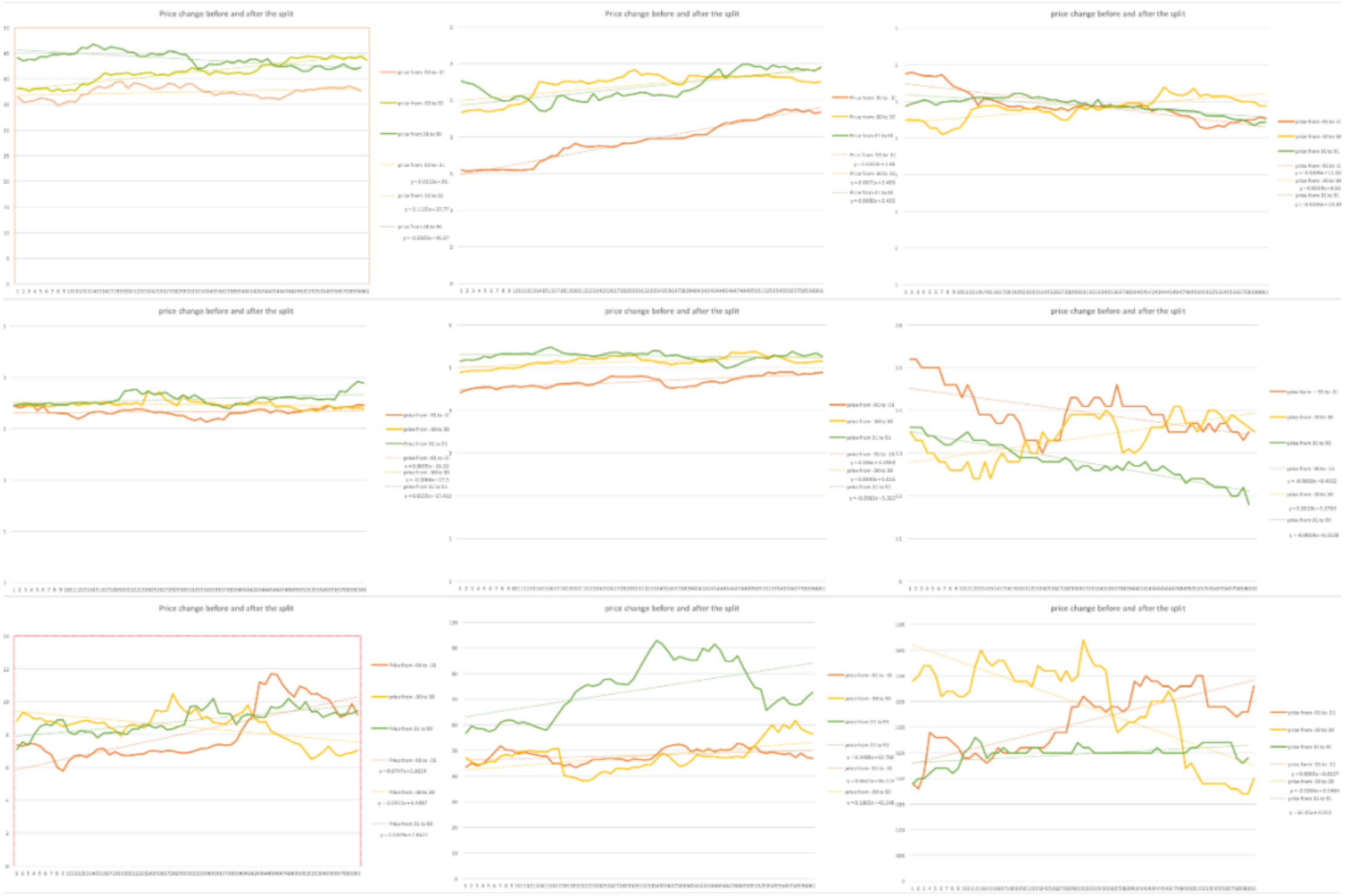

Figure 5. Price change from the day -91 to day +91 across split date. Red lines represent stock price from the day -91 to day -31; Yellow lines represent stock price from the day -30 to day 30; Green lines represent stock price from day 31 to day 91.

Thus, split events appear not to help in reducing the price in order to attract more investors. In fact, investors who sell their shares before the split could gain more profit, than those who sell their shares after the split. What is considered to be worse is that stock splits seem to have an increasing impact on the stock price on average. In Figure 6, the average price from the day -91 to day -31 is the contrast group (blue). It is used to compare with the orange group (the average price from the day -30 day to day 30) and the grey group (the average price from day 31 to day 91). Six orange groups have an increase based on their contrast group; after two months, there still are six grey bars which have an increase in price based on their blue group. According to this, we can deduce that the impact of the stock split in stock price aspect, is apparently not matching manager’s will.

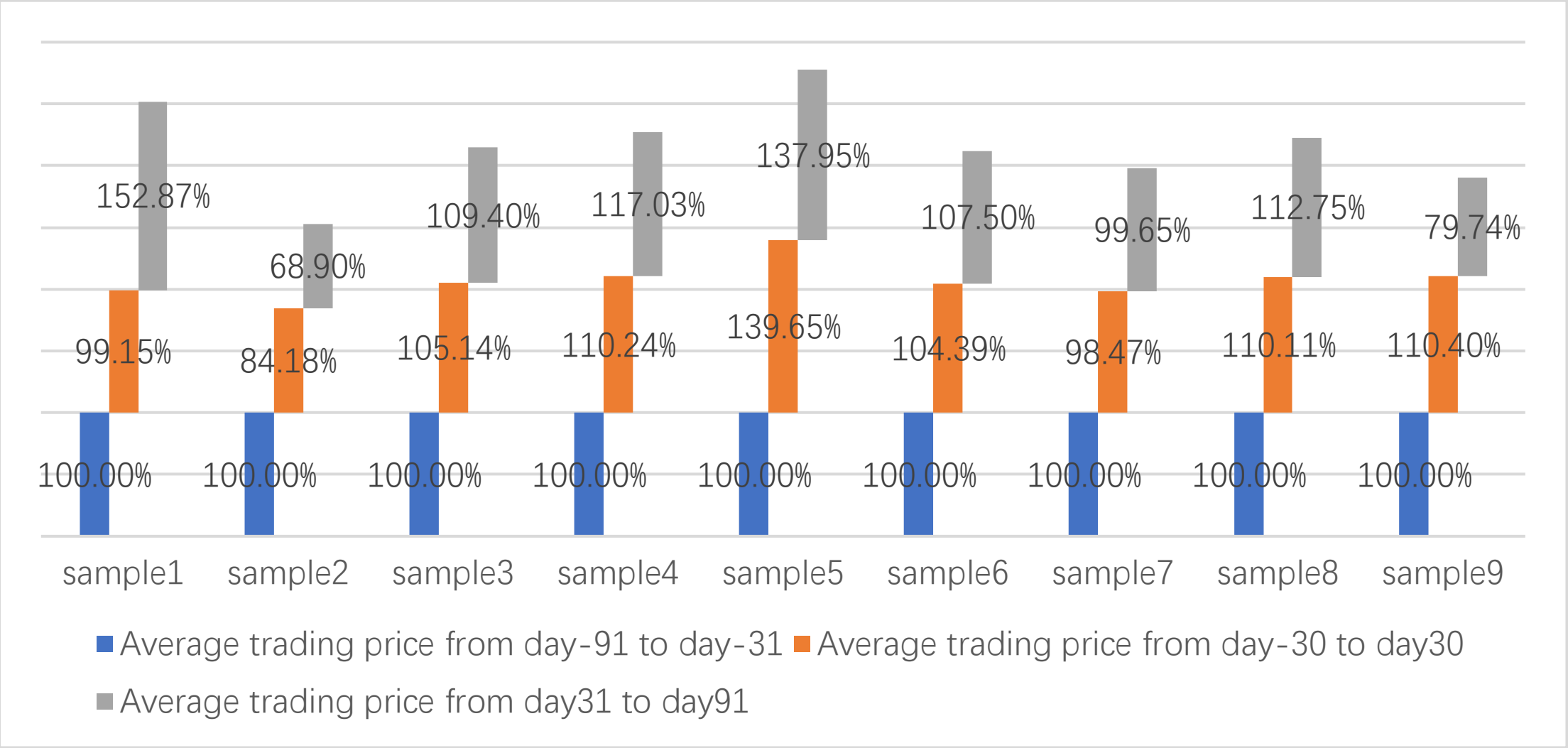

Figure 6. The average price from day -91 to day +91.

Thus, the data analysis so far reports five phenomena:

1. Stock split events have a slight impact on the market.
2. There is an unexpected increase in trading volume for some samples, but “on average”, there is none.
3. There was an explosive demand for shares.
4. Investors are unwilling to trade their shares little by little after the split date.
5. The stock price after the split is higher than the price before the split.

We admit that there is no direct information about the number of shareholders, but we can speculate. After the stock split, the trading number should be higher than that before the split with the split ratio, if the number of shareholders does not change. For example, if the split ratio is 2:1, the trading number after the split date should be double than that before. But, from Figure 2, there is no significant change in trading volumes. Therefore, we can conclude that it is highly likely that the firm increases their shareholder base through stock splits.

### *5.2 Hypothesis 2*

Figure 6 not only induces us into rejecting Baker and Gallagher (1980)~~’s~~ finding but seems to support Karim and Sarkar (2016)~~’s~~ conclusion. In Figure 6, the price after the split is higher, and existed for longer than three months. To ensure that this increase in price did not happen ~~by~~ due to a ‘’panic effect”, let us observe Figure 7.

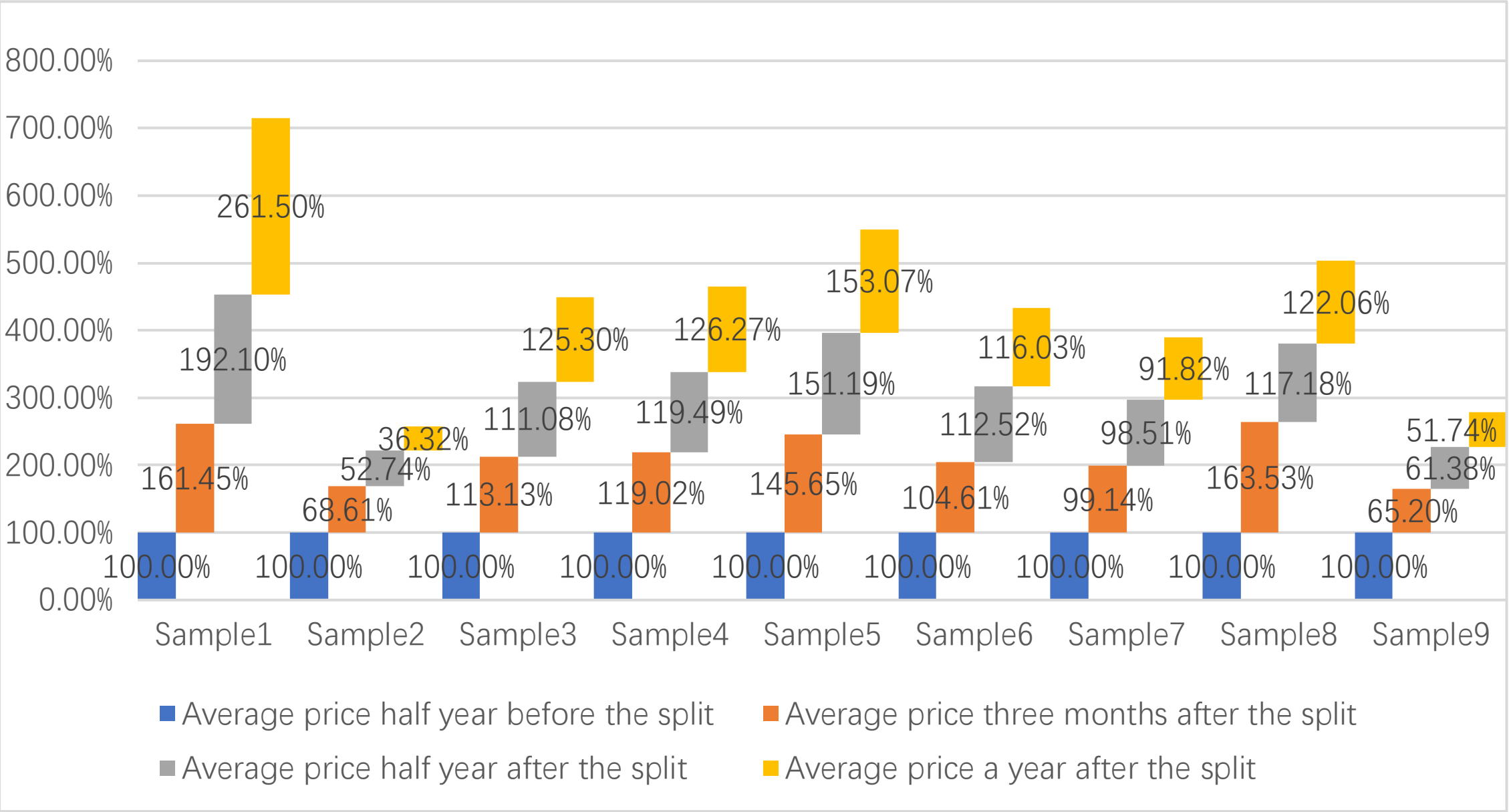

Figure 7. Average stock price for the different periods.

Figure 6 and Figure 7 have a rather similar meaning, but Figure 7 presents advantages in displaying more time periods and longer time intervals. Thus, from Figure 7, one can deduce that most samples have a large price change after three months away from the split date; such trends remain thereafter for one year at least.

In this study, “by split definition”, all samples increased the number of outstanding shares. Therefore, if the value of the firm was priced correctly, then the stock price should have a proportional drop. From Figure 7, one sees that such stock prices violate the market expectation. It mirrors that after revaluation, most of the stocks have a significant change. Six firms had an increase in the price for more than 15%; two drop half of the price, and only sample 7 does not have any significant change.

To check whether the value of sample 2 and sample 9 are unchanged, one can perform a simple calculation. Let

$P_1$ :The stock price before the stock split,

$P_2$ : The stock price after the stock split,

$N_1$ : The number of shares before the stock split,

$N_2$ : The number of shares after the stock split,

$V_1$ : The market value of the sample before the stock split,

$V_2$ : The market value of the sample after the stock split,

whence

$V_1 = P_1 \times N_1$ and $V_2 = P_2 \times N_2$ .

Explicitly, for sample 2, one has:

$N_2 = N_1 \times 1.1$ , $P_2 = P_1 \times 0.52$(after six months) ,

$P_2 = P_1 \times 0.36$(after one year)

$V_2$ (after six months) $= P_1 \times 0.52 \times N_1 \times 1.1 = 0.572\ V_1$

$V_2$ (after one year) $= P_1 \times 0.36 \times N_1 \times 1.1 = 0.396\ V_1$

~~and~~ while for sample 9:

$N_2 = N_1 \times 4.889$ ,

$P_2 = P_1 \times 0.61$(after six months)

$P_2 = P_1 \times 0.51$(after one year)

$V_2$ (after six months) $= P_1 \times 0.61 \times N_1 \times 4.899 = 2.98\ V_1$

$V_2$ (after one year) $= P_1 \times 0.51 \times N_1 \times 4.899 = 2.50\ V_1$

Therefore, one may conclude that the stock split affected all samples' market value.

Furthermore, Figure 7 proves that the increase is not a short-term but rather a long-term effect. In other words, the increase in value is not caused by a panic effect but is due to a fair reassessment by investors, - a behaviour usually attributed to the so called reduction of the information asymmetry. The samples' new value is treated as the benchmark. Figure 7 also provides a substantial evidence supporting Karmin and Sarkar (2016)'s finding, on the percentage of mispriced firms.

Another side of information asymmetries is due to internal information about future earnings. Investors can expect to gain extra future earnings in two ways: increase in the share dividend and increase in the share price. The amount of dividend depends on yearly earnings of the firm and dividend policy. Although the dividend policy for every firm is different, those policies are based on the net profit; whence the condition for shareholders to receive the extra dividend is that firms have increased their net profit. Table 2 provides the percent change in net profit over the studied years for the nine examined samples. The split year net profit is arbitrarily taken as being equal to a 100 (%) value. In Figure 8, one displays the evolution of this net profit for the split year and the subsequent percentage change for the following years. Six samples achieved an increase in the net profit in the following year, but two samples, sample 2 and 3 had a drop. For the examined time intervals, sample 2, 3, and 7 have a decrease in the net profit but recall from Figure 7, that sample 2, 7, and 9 had declined in their stock price one year after the split date. Thus one can observe some dissymmetric evolution: especially, for sample 7 which had a 64.58% increase in net profit in 2015, but its stock price is almost unchanged, whereas sample 3 had a 20% decrease in 2014, but had a 13% increase in the stock price.

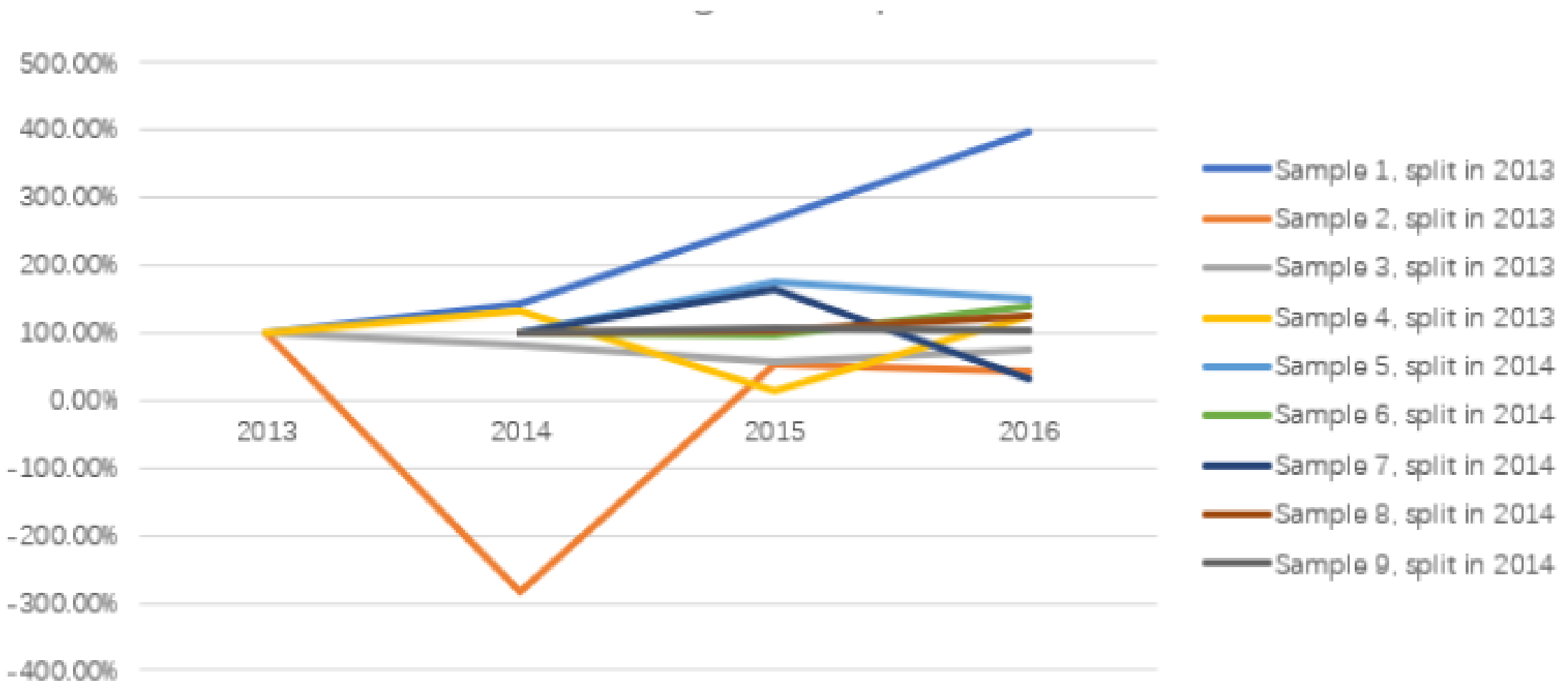

Figure 8. Percentage of change in net profit, for the split year and evolution for the following years.

These features of the sample group again allow us to hypothesize that about one in every five investors has an inverted attitude to the firm value as compared with the net profit change. Thus, on one hand, the data contradicts the findings by Brennan and Copeland (1988) over investors reaction at split times; on the other hand, the findings provide a measure for the number of ‘’irrational agents” as discussed in Dhesi & Ausloos (2016).

Table 2. Percent change in net profit over the studied years for the nine examined samples; the split year is mentioned in parentheses in the first column. The split year net profit is arbitrarily taken as being equal to a 100 value.

| % change in net profit | 2013 | 2014 | 2015 | 2016 | | Total Diff. |
|---|---|---|---|---|---|---|
| #1(2013) | 100 | 143.94 | 268.58 | 399.29 | | 299.29 |
| #2(2013) | 100 | -282.94 | 54.89 | 40.92 | | -59.08 |
| #3(2013) | 100 | 80.65 | 58.17 | 73.96 | | -26.04 |
| #4(2013) | 100 | 130.54 | 15.40 | 126.35 | | 26.35 |
| | | | | | | |
| #5(2014) | | 100 | 176.97 | 150.26 | | 50.26 |
| #6(2014) | | 100 | 96.48 | 140.17 | | 40.17 |
| #7(2014) | | 100 | 164.58 | 31.54 | | -68.46 |
| #8(2014) | | 100 | 104.01 | 124.48 | | 24.48 |
| #9(2014) | | 100 | 107.34 | 103.91 | | 3.91 |

The ROE (return on Equity) ratio provides a different angle for examining this hypothesis. A ROE ratio is the amount of the net income returned as a percentage of shareholders’ equity; it further relates the net profit with shareholders’ investment. The ROE ratio is a measure of how much profit a company generates with the money shareholders have invested and is an indicator of how well the firm is optimizing shareholders’ wealth. Investors who invest their money into a firm are wishing and believing that the firm could efficiently use their investment. However, the table in Figure 9 shows a disappointing result. Two years after the operation, no samples had a significant improvement. The best performance is for sample 1 and sample 2; both improved their efficiency in equity by 9% only.

The finding is a surprise: indeed, sample 2 had the largest drop in the yearly net profit and its stock price, but sample 2 also had maximal improvement in its efficiency of maximising shareholders' wealth. From Figures 7-9, five samples (and see Table 3) have the same changing trend in their stock prices, the yearly net profit and return on equity ratio. Those five samples were more likely mispriced by the market. Thus, the stock split announcement has interestingly helped the market into fairly priced them.

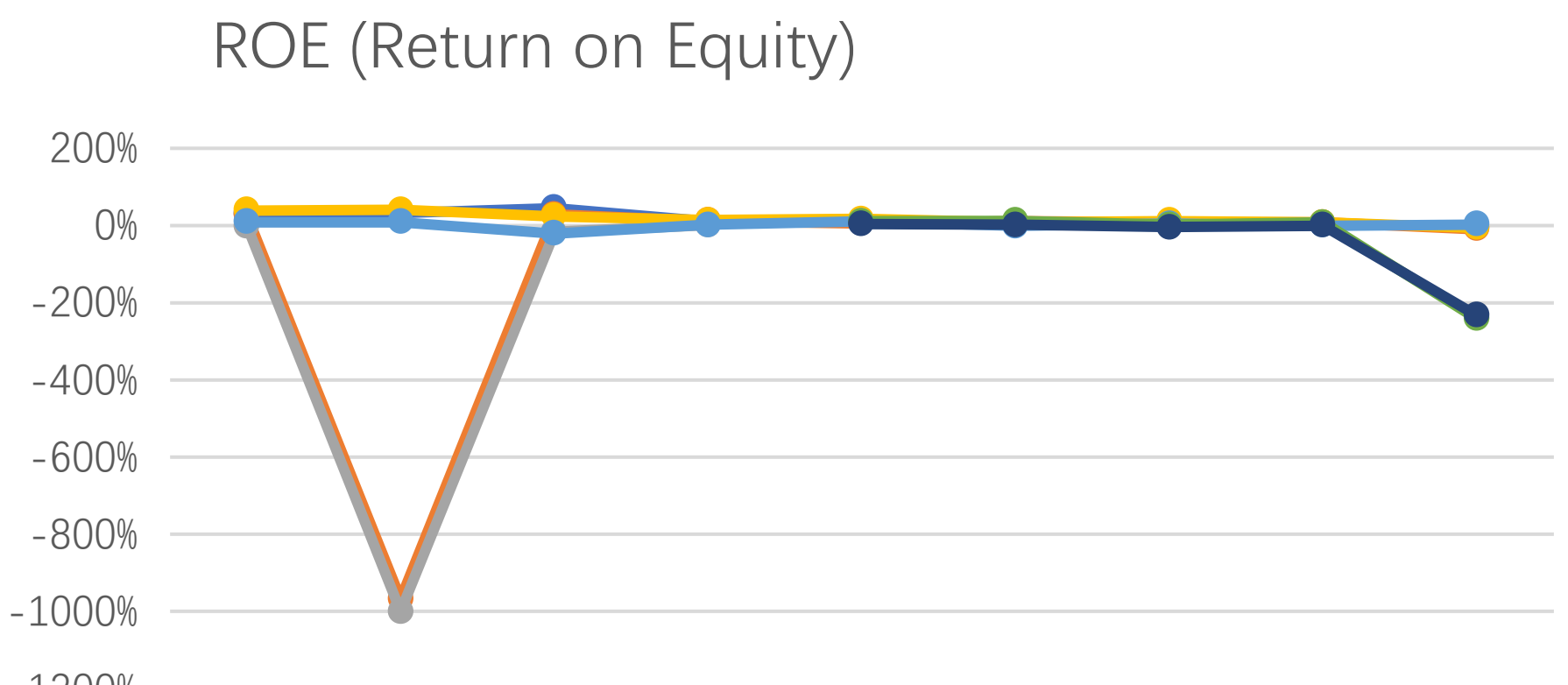

| | sample 1, split in 2013 | sample 2, split in 2013 | sample 3, split in 2013 | sample 4, split in 2013 | sample 5, split in 2014 | sample 6, split in 2014 | sample 7, split in 2014 | sample 8, split in 2014 | sample 9, split in 2014 |
|---|---|---|---|---|---|---|---|---|---|
| 2013 | 30% | 32% | 45% | 11% | | | | | |
| 2014 | 28% | -967% | 29% | 12% | 6% | 10% | 7% | 7% | -9% |
| Percertage change 13-14 | -2% | -999% | -16% | 1% | | | | | |
| 2015 | 39% | 41% | 24% | 13% | 17% | 9% | 12% | 7% | -6% |
| Percertage change 13-15 | 9% | 9% | -21% | 2% | 11% | -1% | 5% | 0% | 3% |
| 2016 | | | | | 10% | 12% | 2% | 7% | -242% |
| Percertage change in 14-16 | | | | | 4% | 2% | -5% | 0% | -233% |

Figure 9. ROE ratio from 2013 to 2014.

Table 3. Samples with positive ROE.

| | Sample1 | Sample4 | Sample5 | Sample6 | Sample8 |
|---|---|---|---|---|---|
| The percentage of increase in the stock price | 165.5% | 26.27% | 53.07% | 16.03% | 22.06% |
| The percentage of increase in the net profit | 299.29% | 26.35% | 50.26% | 40.17% | 24.48% |
| The percentage of increase in the return on equity | 9% | 2% | 4% | 2% | 0% |

The way for investors to receive earnings is when selling shares. The profit from selling is a relay on the price difference. According to Figure 10, sample 1, 3, and 5 had a high percentage increased in the stock price for the eight months. Investing in those three samples could lead to expect a better profit than an investment in the other samples. However, Figure 11 shows a different outcome.

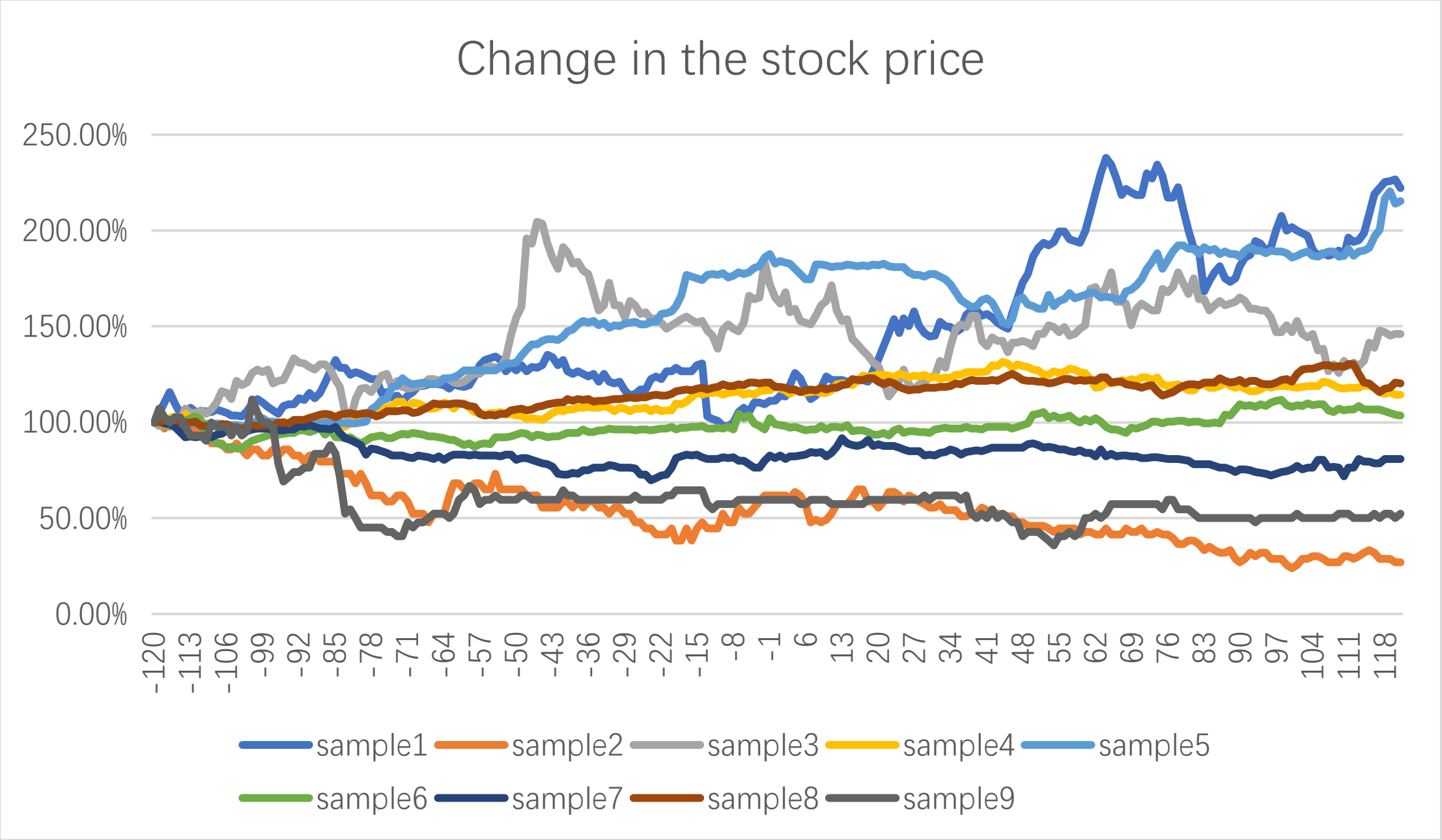

Figure 10. Change in the stock price four months before and after the split date.

- After the first month, investors of sample 1, sample 4, and sample 7 could receive 30.59%, 3.34% and 2.69% abnormal earning separately.
- After two months, investors of sample1 (63.53%), sample 3 (0.35%), sample 4 (0.01%), sample 6(7.45%), and sample 8 (3.29%) could gain abnormal earnings.
- After a season, investors of sample 1(102.21%), sample 5 (9.59%), sample 6 (7.45%), and sample 8 (2.36%) could obtain an abnormal return.
- After four months, investors of sample 1 (72.94%), sample 3 (13.68%), sample 5 (23.41%), sample 6 (11.66%), and sample 8 (1.87%) will get the extra earnings.

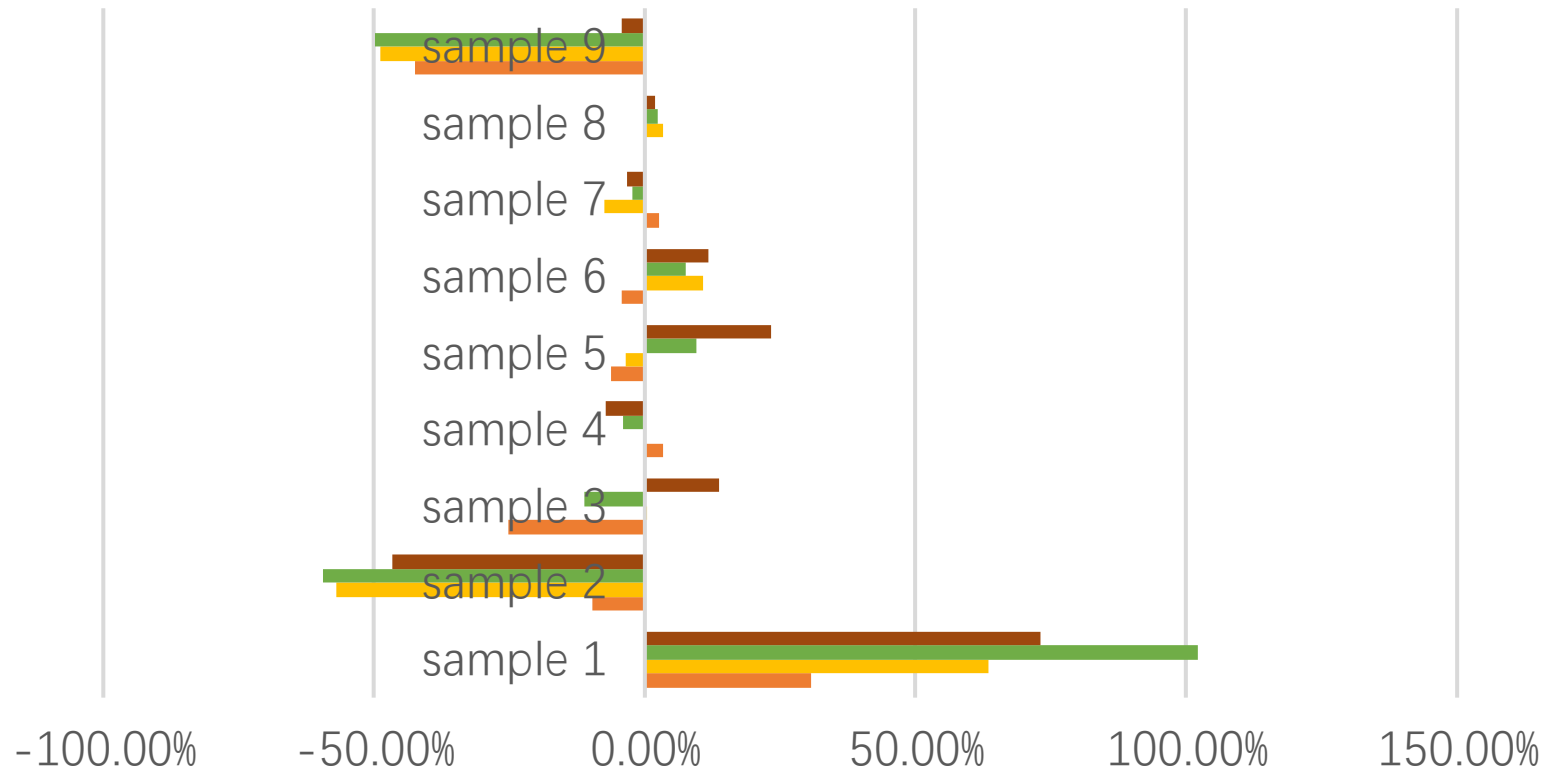

| | sample 1 | sample 2 | sample 3 | sample 4 | sample 5 | sample 6 | sample 7 | sample 8 | sample 9 |
|---|---|---|---|---|---|---|---|---|---|
| abnormal return after four months | 72.94% | -46.65% | 13.68% | -7.18% | 23.41% | 11.66% | -3.32% | 1.87% | -4.25% |
| abnormal return after three months | 102.21% | -59.38% | -11.15% | -4.11% | 9.59% | 7.45% | -2.24% | 2.36% | -49.71% |
| abnormal return after two months | 63.53% | -56.90% | 0.35% | 0.01% | -3.49% | 10.78% | -7.39% | 3.29% | -48.97% |
| abnormal return after one month | 30.59% | -9.76% | -25.32% | 3.34% | -6.28% | -4.23% | 2.69% | -0.23% | -42.56% |

Figure 11. Abnormal return rates.

Figure 10 shows the actual amount that the investors could gain from trading shares while Figure 11 shows the expectation that investors want to make from trading shares. For example, sample 6, in Figure 10, shows that the stock price of sample 6 is not as high as the price of sample 3, but the stock price performance of sample 6 is higher than investors' expectation. Furthermore, investors of sample 6 had a higher level of misunderstanding to the value of the firm, - or sample 6 had more covered information than sample 3. But, after the split date, the samples have a positive abnormal return, which means that the stock market generally operates better than investors expected. Recall that Brennan and Copeland (1988) found that investors are readjusting their beliefs about the value of the firm. If investors correctly readjusted their beliefs, the "abnormal return" should be equal to zero. However, data from Figure 12 suggest that investors do not readjust perfectly; most investors had either over or under expectation about their investment. The reduction in information asymmetry does not necessarily lead to a uniform herding behaviour! This leads to unexpected tail exponents in the return statistical distributions (Dhesi & Ausloos, 2016).

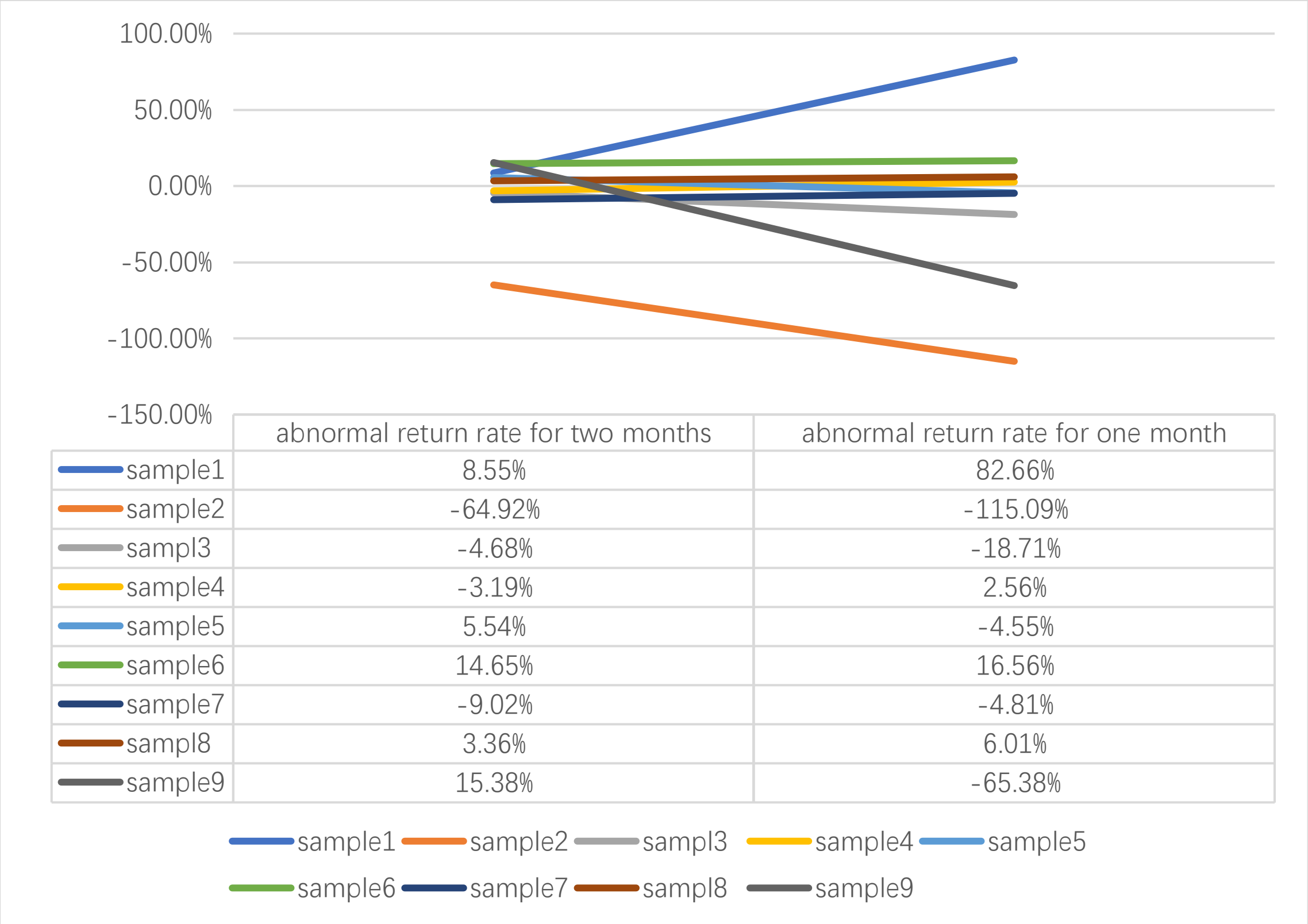

Figure 12. Abnormal return rate with virtual demarcation day.

Table 3. Beta of each sample.

| Sample1 | Sample2 | Sample3 | Sample4 | Sample5 | Sample6 | Sample7 | Sample8 | Sample9 |
|---|---|---|---|---|---|---|---|---|
| -0.09066 | 0.04821 | 0.01182 | -0.01784 | -0.16377 | 0.01110 | 0.08801 | 0.04164 | -0.00246 |

## 5.3 *Hypothesis 3*

From McNichols and Dravid's review (1990), one deduces that a lower range of share prices enhances the liquidity of a firm's shares. Figure 13 shows the price gap 90 days before and 90 days after the split date, - blue and yellow lines respectively. Either from the average level or the number of high points, it can be observed that yellow lines are below blue lines. This indicates a good result for the firm, because a low price gap will attract investors' consideration, be attractive to analysts for reassessment of its value, and then affects its brand value for the public. Thus, as one of the indicators to detect the change of the liquidity, the price gap as displayed on Figure 13, in general, provides evidence of increasing liquidity for the 9 cases.

The second element to be considered is the trading volume (Lo & Wang, 2000). In the market with poor liquidity, there are fewer trades executed every day. Under an extreme assumption, if there is a stock market with only one executive order every day, then for this stock market, the price gap is zero whenever investors test it. Hence, the trading volume, as previously displayed in Fig. 3, is notably also another indicator to examine the liquidity.

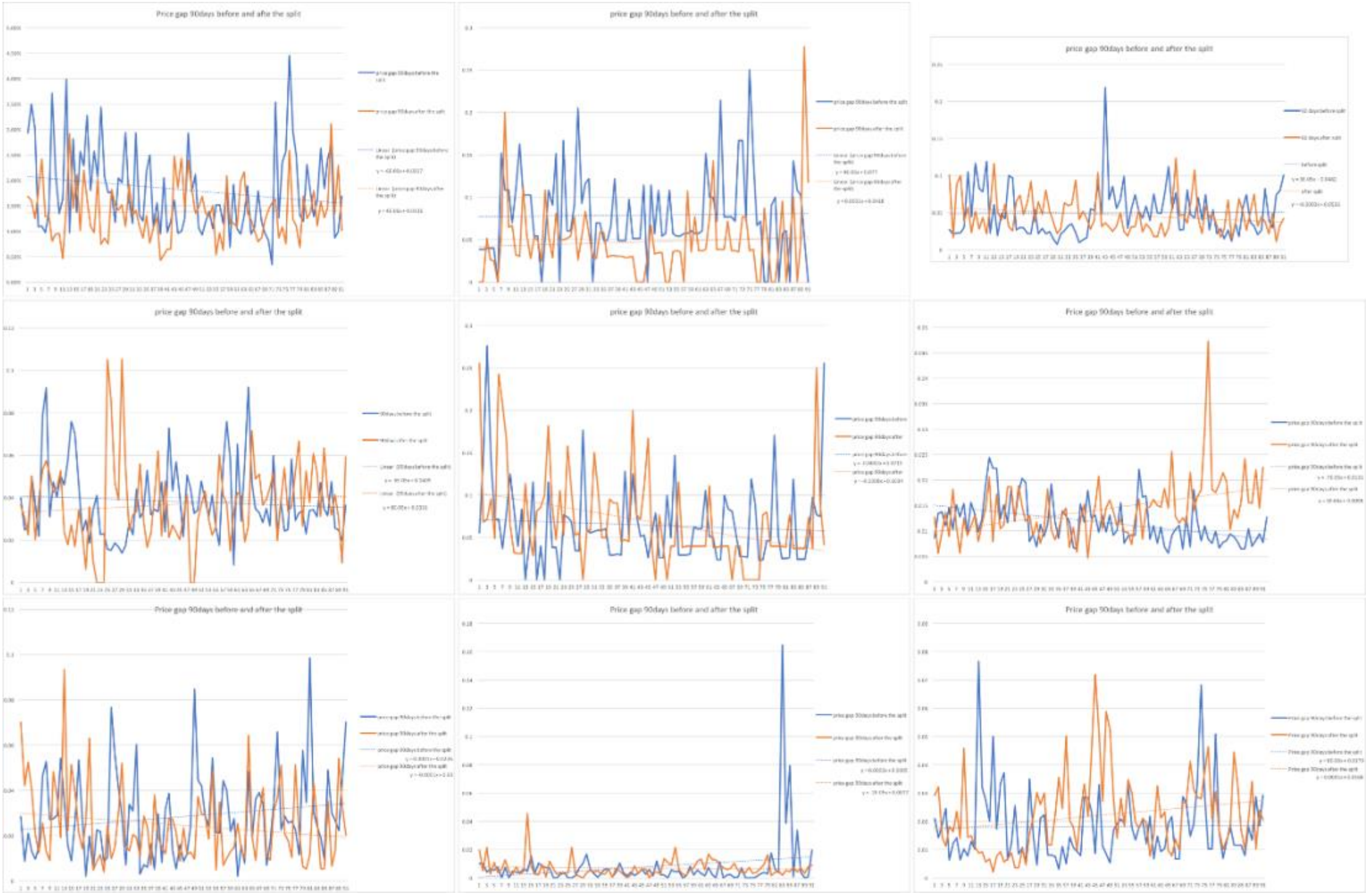

Figure 13. Price gap over 90 days before and after the split date. The blue line present the price gap 90 days before the split and yellow line represent the price gap 90 days after the split.

Recall that Figure 3 shows the daily trading volume 30 days before and 30 days after the split day. Combined with Figure 2 and Figure 4, those three Figures point to the effect of the stock split in the short-term; they indicate that after the stock split, the trading volume is slightly affected, but its effect is falling day by day. Because time periods in Figure 13 and Figure 3 are different, the conclusion from Figure 3 can only be indirectly used in analysing Figure 13. In fact, Figure 14 shows that the trading volumes for most stocks increased. The outcome of Figure 14 is different from that of Figure 3 but supporting Figure 13. Over a long time span, the stock split had improved effect to the market in trading amount and also reduced the price gap. These advantages further reflect the rising liquidity of the market.

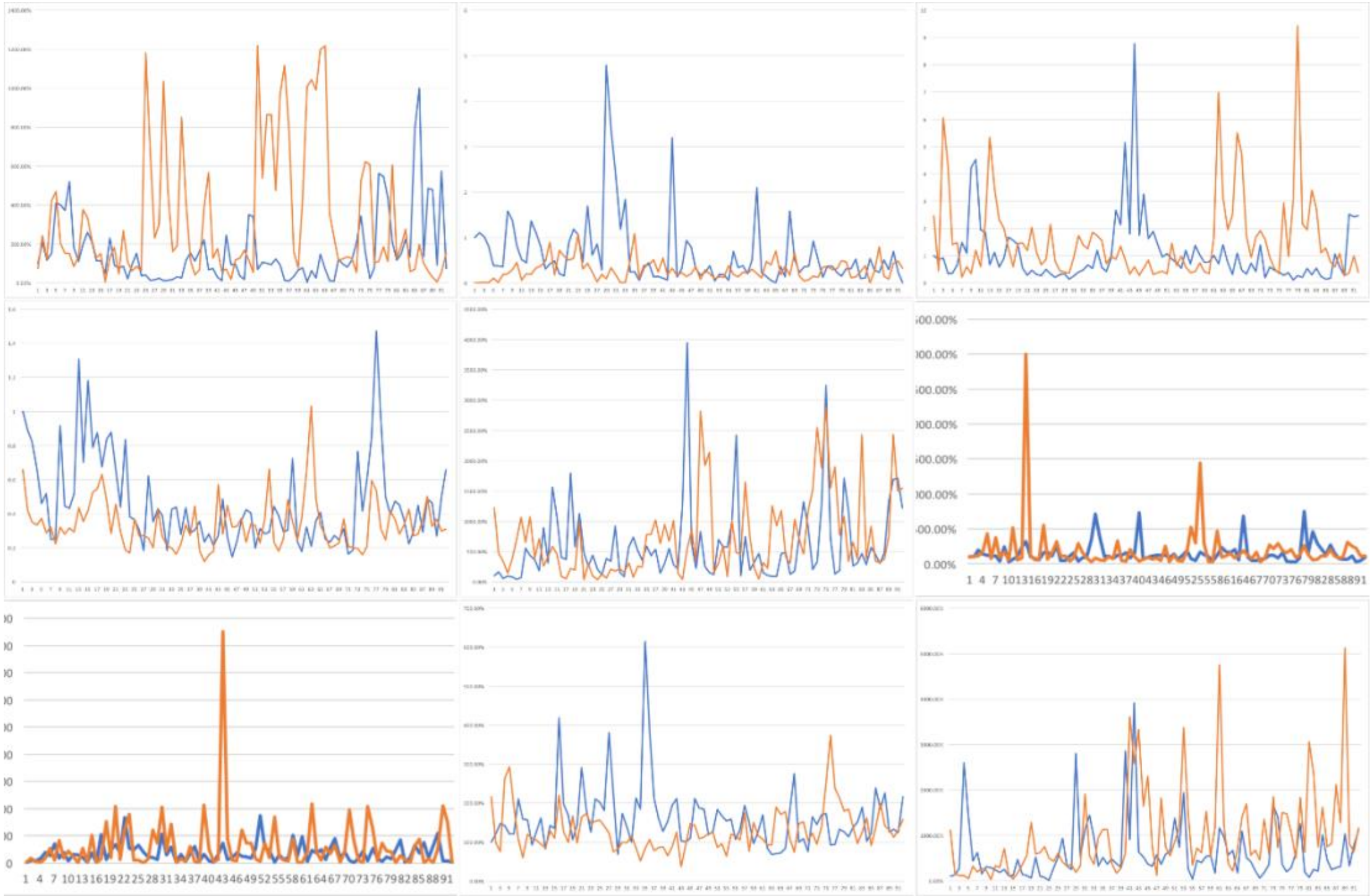

Figure 14. The daily trading volume in 90 days before and after the split, the blue line represents the daily trading volume during 90 days before the split and the yellow line the daily trading volume during 90 days after the split.

The differences between Figure 3 and Figure 14 point to an important element in the reasoning: the time interval size. Different time intervals apparently lead to different conclusions; in brief, a long time interval appears to emphasize the influence of the stock split in a more convincing way.

In Figure 15, we show in yellow the price gap three months after the split and six months before the spit.. In the blue part, the price gap of five stocks increased. At the end of three months, three of nine samples had an increasing trend in price gap. To some of the stocks, the repercussion in price gap happens after half a year, but for most of the stocks, the decline in price gap remains for a more extended period.

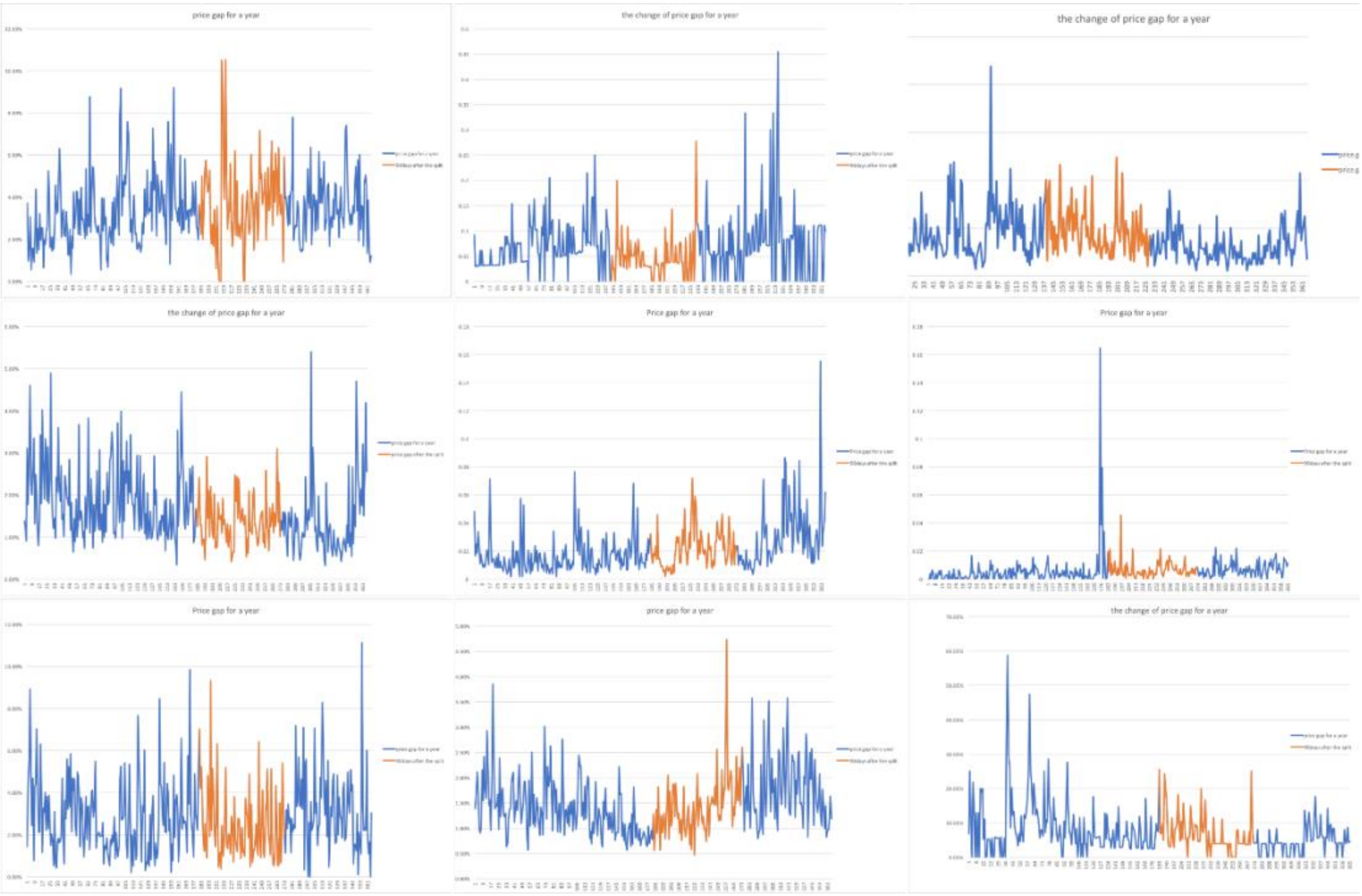

Figure 15. The price gap half a year before and after the split day: the blue line represents the price gap for the year and the yellow line represents the price gap during 90 days after the split day.

In Figure 16, we show that the after split daily trading amount is on average higher than that before the split, although there is no significant signal to point out toward a high trading volume with low price gap; those three stocks which have a rising trend in their price gap also have a diminishing trend in their trading volume. From this phenomenon, we propose to link the change in price gap with the shift in trading volume together and infer that those two changes have an inversely proportional relationship. Moreover, from this inference, we can state that an increase in trading volume corresponds to a decline in price gap.

Moreover, those two phenomena together can generate some improvement in the liquidity of the stock market. Data from Figure 13 to Figure 16 is testifying for this finding.

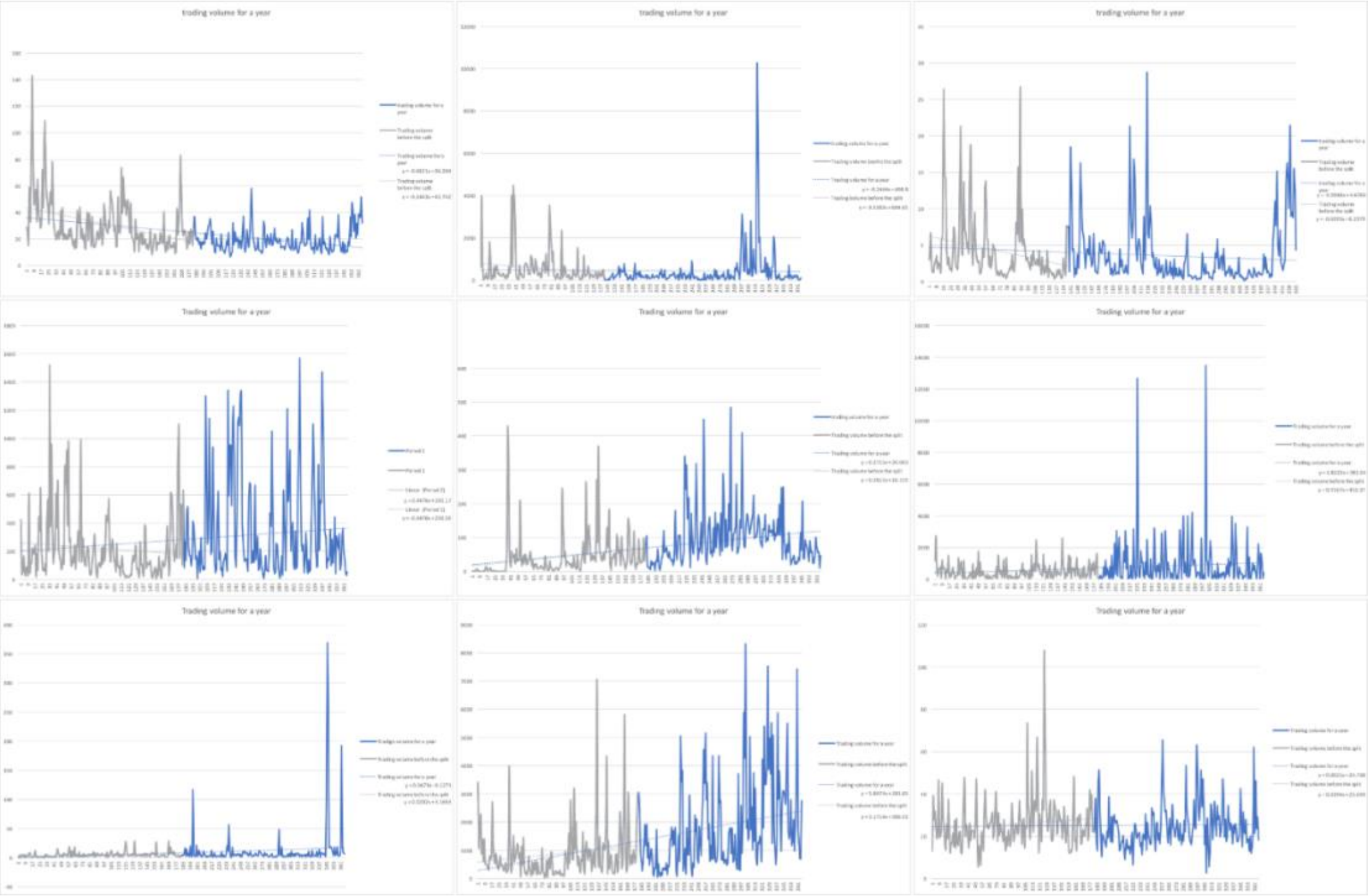

Figure 16:. Trading volume half a year before and after the split day. The blue line represents the trading volume half year after the split day, and the grey line represents the trading volume half year before the split.

## 6. Conclusion

This paper reports studies on stock split events and tries to propose answers to the often raised questions “why do companies split their stock?”, and ‘’why do shareholders often agree on such splits?”

Theoretically, every shareholder receives a number of shares according to a split ratio after the stock split date. Yet, the market price per share necessarily drops according to the split rate to keep the firm's market value approximately constant. However, in practice, it is rare to see the price reduced by half for a 2:1 stock split. For example, in Figure 5, the price of the stock has a slight fluctuation only after the split date; generally, there is no significantly drastic increase or drop.

Therefore, we contribute to the field knowledge by observing these features:

1. Stock splits affect the market and slightly enhance the trading volume in a short-term.
2. Stock splits increase the shareholder base for the firm.
3. Most of the firms are mispriced in the split year; stock split announcements reduce the level of information asymmetries.
4. Investors readjust their beliefs to the firm, but (unfortunately?) most of the investors still hold an inexact "fundamental image" of the firm.
5. Split shares have a positive effect on the liquidity for the market.

Moreover, there are at least advantages for firms to split shares, so called strategy benefit. The stock split is a proper way to readjust the firm's value, compatibility, and marketability.

Stock splits do present some, but not perfect, informational content (Grinblatt et al., 1984). However, for shareholders, stock split announcements are no more than an action taken by the fir; cruelly to say, there is no real information. All that information is given by generally accepted, implied rules. Bypassing those rules, a forthcoming share split announcement seems to amount to nothing more than paperwork.

Data and analysis in the present study also show that not every stock splits lead to the same result. There are some firms which after a stock split went bankrupt after few years. Therefore, the information which is passed by the split decision is a theoretical prediction for the future but not a practical prophecy for bettering.

Thus, managers who apparently feel that unless stock splits are viewed as an occasional device, in which the impact of the division would be fully reflected in the stock's price, are somewhat fallacious reasoners like sophists (Baker & Gallagher, 1980).

In conclusion, every stock split is an unreliable strategic event for the public, even though, as we have mentioned, there are many studies about the stock split and hypotheses about influences of the stock split. With such uncertain futures, a split plan appears like a mist; any change in the environment, even a little one, could make split plan nondeterministic.

We should pose in pointing out that we are aware of several limitations in our study. In some sense this allows us to consider that how pleased we would be, with others, to see that this study is expanded. We studied only 9 cases among the 7727 events in 2013 and 2014; no need to say that more cases, some selected in a non-random way, thus considering types of businesses, fields of activities, and other distinguishing parameters would be of interest. There may be more internal factors that influence the stock split events, like ''Corporate Social Responsibility" (Harjoto, Kim, Laksmana, & Walton, 2019). In fact, external factors might also be relevant, in particular in the most recent times. Correlations between effects of the COVID-19 pandemic (Afdhal, Mayapada, & Septian, 2022) and the subsequent global crisis, sustainability, Sustainable Development Goals, and nowadays the Ukrainian war, seem to be modern variables to consider. Stock split cases in many other countries, beside UK, have already been examined, (see recent references in Burnwal & Rakshit (2018)), but are to be examined, - again within different time intervals and various variables. Beside countries, one can focus on digital platforms and the improvement of learning outcomes, based on the evidence extracted from meta-analysis; these are key focuses nowadays. See in this journal recent work by Popescu and Popescu (2019) on the matter.

Acknowledgements

We thank the editor and reviewers for their constructive criticism.

Authors Contributions:

Conceptualization, JNC and MA; Methodology, JNC and MA; Software, JNC and MA; Validation, JNC and MA; Formal Analysis, JNC and MA; Investigation, JNC and MA; Data Curation, JNC and MA; Writing-Original Draft Preparation, JNC and MA; Writing-Review & Editing, JNC and MA; Visualization, JNC and MA; Supervision, MA. The authors contributed to all the various aspects of this research and paper production within their respective skill and expertise. The final form of this manuscript was read and approved for publication by the authors; the work described represents an original research that has not been published previously and is not under consideration for publication elsewhere, in whole or in part.

Conflict of interest statement

The authors declare that the research was conducted in the absence of any philosophical, commercial or financial relationships that could be construed as a potential conflict of interest.

Data availability statement :

The analyzed data is freely available on internet and can be obtained through websites mentioned in the text and in the Bibliography.